 \documentclass[twocol]{ametsoc}

\usepackage{lscape} 
\usepackage{rotating}
\usepackage{longtable} 
\usepackage{color,soul} 
\usepackage{url}

\newcommand{\about}{\mathord{\sim}}

\xdef\currjournal{}
\runningheadauthors{Rasmussen et al.}
\bibpunct{(}{)}{;}{a}{}{,}

\title{Probability-weighted ensembles of U.S. county-level climate projections for climate risk analysis}

    \authors{D. J. Rasmussen}
    \affiliation{Rhodium Group, Oakland, CA, USA \thanks{Now at:  Woodrow Wilson School of Public \& International Affairs, Princeton University, Princeton, NJ 08540, USA}}
    
    \extraauthor{Malte Meinshausen}
    \extraaffil{Australian-German Climate \& Energy College, School of Earth Sciences, The University of Melbourne, Parkville 3010, Victoria, Australia \\ 
                    The Potsdam Institute for Climate Impact Research, Telegraphenberg A26, 14412 Potsdam, Germany}

    \extraauthor{Robert E. Kopp\correspondingauthor{Department of Earth \& Planetary Sciences, 610 Taylor Road, Rutgers University, Piscataway, NJ 08854, USA}}
    \extraaffil{Department of Earth \& Planetary Sciences, Rutgers Energy Institute, and Institute of Earth, Ocean \& Atmospheric Sciences, Rutgers University, New Brunswick, NJ, USA}
    \email{robert.kopp@rutgers.edu}
        
\abstract{Quantitative assessment of climate change risk requires a method for constructing probabilistic time series of changes in physical climate parameters. Here, we develop two such methods, Surrogate/Model Mixed Ensemble (SMME) and Monte Carlo Pattern/Residual (MCPR), and apply them to construct joint probability density functions (PDFs) of temperature and precipitation change over the 21st century for every county in the United States. Both methods produce \emph{likely} (67\% probability) temperature and precipitation projections consistent with the Intergovernmental Panel on Climate Change's interpretation of an equal-weighted Coupled Model Intercomparison Project 5 (CMIP5) ensemble, but also provide full PDFs that include tail estimates. For example, both methods indicate that, under representative concentration pathway (RCP) 8.5, there is a 5\% chance that the contiguous United States could warm by at least 8$^\circ$C. Variance decomposition of SMME and MCPR projections indicate that background variability dominates uncertainty in the early 21st century, while forcing-driven changes emerge in the second half of the 21st century. By separating CMIP5 projections into unforced and forced components using linear regression, these methods generate estimates of unforced variability from existing CMIP5 projections without requiring the computationally expensive use of multiple realizations of a single GCM.} 

\begin{document}

\maketitle

%

\section{Introduction}

The risk of an adverse event is characterized by its probability and its consequences \citep{Kaplan1981a}. Risk analysis thus requires consideration of the probabilities and consequences of as full a range of possible outcomes as possible, including `tail risks' that are low probability but high consequence.

For assessments of the local and regional risks of climate change, this poses two major challenges. First, ensembles of coupled atmosphere-ocean general circulation models (GCMs) and Earth system models (ESMs), such as those in the Coupled Model Intercomparison Project Phase 5 (CMIP5) archive \citep{Taylor2012a}, are not probability distributions and were not designed to consider all sources of projection uncertainty. CMIP5 model ensembles are ``ensembles of opportunity'', arbitrarily compiled on the basis of modeling center participation. Sampling from such a distribution by assigning equal probability to all models may therefore yield a biased outcome \citep{Tebaldi2007a}. Second, GCMs and ESMs may underestimate the probability of extreme climate outcomes. For example, the range of the equilibrium climate sensitivity (ECS) in the CMIP5 is  2.1--4.7$^\circ$C per doubling of CO$_2$ concentrations \citep[see][table 9.5]{Flato2013a}, while observational and other non-GCM constraints allow a $\about 17\%$ probability of values exceeding 4.5$^\circ$C \citep{Collins2013a}. Simply weighting individual GCMs in a multi-model ensemble will not produce such extreme behavior if it is not simulated. Quantitative risk analysis that leverages the detailed physical projections produced by GCMs therefore requires methods that (1) assign probability weights to projected changes and (2) account for tail risks not captured by the physical models.

In this study, we develop two such methods and demonstrate them by producing county-level projections of 21st century changes in temperature and precipitation in the United States. The first method, Surrogate/Model Mixed Ensemble (SMME), uses probabilistic simple climate model (SCM) projections of global mean temperature change to weight GCM output and to inform the construction of model surrogates to cover the tails of the SCM probability distribution that are missing from the GCM ensemble. The second method, Monte Carlo Pattern/Residual (MCPR), decomposes GCM output into forced climate change and unforced climate variability, uses SCM temperature projections to scale patterns of forced change, and then adds unforced variability. The SMME projections presented here were recently applied in a quantitative analysis of some of the economic risks climate change poses to the United States (U.S.) \citep{Houser2015a}.

While perturbed physics ensembles \citep[e.g.,][]{Stainforth2005a} can produce PDFs of future climate through sampling projection uncertainty originating from model parameters, this approach requires enormous computing resources. However, SCMs [e.g, the Model for the Assessment of Greenhouse Gas-Induced Climate Change (MAGICC; \citet[]{Meinshausen2011b}] can be run in a probabilistic fashion on a desktop computer, sampling the range of parametric uncertainty consistent with both historical observations and expert judgement of parameters such as climate sensitivity. In addition, MAGICC has shown to well emulate global mean temperature from GCMs over multiple emissions scenarios \citep[e.g.,][]{Rogelj2012a}, ensuring that SCM generated PDFs encompass both the spread of results from key variables in the CMIP5 archive and global mean temperature pathways not simulated in complex models.

Model surrogates used to cover the tails of the PDF must spatially resolve local projections of climate change under global temperature pathways not present in GCMs. Pattern scaling applies a linear relationship between changes in local climate variables and coincident changes in global temperature (i.e., patterns) produced by GCMs with a scalar (time-evolving global mean temperature) to generate projections under alternative global temperature pathways that would otherwise require a GCM to simulate \citep{Santer1990a,Mitchell2003a,Moss2010a}. Moreover, the same linear regression used for pattern scaling can facilitate uncertainty quantification. If projections from a GCM are considered as the sum of forced and unforced climate variability, linear regression can disentangle these components with the forced signal estimated as the linear trend and the residuals representing a first-order approximation of unforced variability. While conventional pattern scaling approaches discard the latter, as they are uncorrelated with global mean temperature, we retain these to assess the projection uncertainty associated with unforced variability and to compare with estimates from computationally expensive multi-member initial condition ensembles \citep[e.g.,][]{Kay2014a,Deser2014a}.

In Section 2 of this paper, we first present an \emph{a priori} comparison of the approaches and then detail the methodologies. In Section 3, we identify sources of agreement and disagreement between temperature and precipitation results from an equal-weighted GCM ensemble, SMME projections, and MCPR projections and  examine their uncertainties. In Section 4, we consider the implications of these comparisons for the application of the two methods. We summarize the main findings and conclude in Section 5. Additional tables, figures, and further detailed methodology appear in the appendices. All daily projections compiled in this analysis are freely available at: [URL TBD on Rutgers library archive]. 

\section{Methods}

General overviews of both probabilistic methodologies are shown in Fig. \ref{fig:flow}. In each case, we start with an estimated probability distribution of global mean temperatures over time from a SCM. For the SMME method (Fig. \ref{fig:flow}A), we use SCM projections of temperature change over the 21st century to weight GCM projections of monthly temperature and precipitation that have been bias-corrected and downscaled using the bias-corrected spatial disaggregation (BCSD) method \citep{Brekke2013a} (see Appendix A) and `surrogate' models employed to ensure the tails of the probability distribution are represented. For the MCPR method (Fig. \ref{fig:flow}B), the pathways of temperature change projected by the SCM are combined with randomly selected patterns of forced change and residuals of unforced variability from the downscaled CMIP5 models. 

\subsubsection{A priori comparison of the approaches}

We note three potentially important differences between the SMME and MCPR approaches.

First, within the range of global temperatures for which CMIP5 output is available, the SMME approach allows for more complex, non-linear relationships between global temperature and regional forced change than the MCPR method, which assumes a constant relationship reflected by the patterns. 

Second, the patterns and residuals employed in the two approaches are selected differently. The SMME method requires \emph{ad hoc} selection of the patterns used to create surrogate models, while the MCPR applies a consistent algorithmic method to generate all output. Furthermore, while the SMME method retains a pairing between patterns and residual, the MCPR approach assumes that patterns and residuals are independent of one another, which is unlikely to be strictly true. Feedbacks between both components are possible. For instance, the external forcing may affect the properties of the background variability, such as its variance. As used here, the MCPR method assumes all patterns and residuals are equally likely. In the SMME technique, the patterns and residuals of models associated with higher-probability global temperature projections have greater weight.

Third, the SMME method uses SCM global mean temperature change in 2080--2099 as the target for the probability distribution, but may deviate from the SCM distribution at other time points. For long-term, global mean temperature change, the MCPR approach will always match the SCM distribution, as all patterns perfectly track a specific quantile of SCM global mean temperature.

\subsection{Global mean temperature}
\label{globalmeantemperature}

We incorporate radiative forcing projections from all four representative concentration pathways (RCPs): RCP 2.6, RCP 4.5, RCP 6.0, and RCP 8.5 \citep{VanVuuren2011a}. Projections of global mean temperature for the four RCPs are calculated using MAGICC6 \citep{Meinshausen2011b} in probabilistic mode. MAGICC6 is an SCM that represents hemispherically-averaged atmosphere and ocean temperature and the globally-averaged carbon cycle. MAGICC6 does not simulate internal climate variability or precipitation, both of which require more complex models. The distribution of input parameters for MAGICC6 that we employ has been constructed from a Bayesian analysis based upon historical observations of hemispheric land and ocean surface air temperature, ocean heat content, estimates of radiative forcing \citep{Meinshausen2009a}, and the ECS probability distribution from AR5 \citep{Collins2013a} (Fig. A1). The climate sensitivity probability distribution from AR5 is based on several lines of information. Observational, paleoclimatic, and feedback analysis evidence indicate 5th/17th/83rd percentiles of 1.0$^\circ$C/1.5$^\circ$C/4.5$^\circ$C. Additional evidence from climate models suggests a 90th percentile of 6.0$^\circ$C. The differences in climate sensitivity between MAGICC6 and AR5 in part reflect sampling and the constraints needed to fit historical observations within the MAGICC6 model structure.

\subsection{Pattern fitting}
\label{patternscaling}

Assuming that forced climate change can be approximated as linear in the long-term (30-year) running average of global mean temperature, for each CMIP5 model and scenario $i$ and each at station $j$, we fit the changes from the 1981--2010 reference levels for seasonal temperature and precipitation to the linear model
\begin{eqnarray}
y_{i,j}(\Delta T,t) = \hat{k}_{i,j} \Delta T + b_{i,j} + \epsilon_{i,j}(t) \label{eqn:patternscaling}
\end{eqnarray}
following \citet{Mitchell2003a}. Here, $\Delta T$ is the running-average change in global mean temperature relative to the reference period (1981--2010), $\hat{k}$ is the estimated seasonal pattern, $\hat{k} \Delta T$ is the estimated forced climate change, $b_{i,j}$ is the observed historical mean, and $\epsilon(t)$ is an estimated temporal pattern of unforced variability. As an example, Fig. A2 shows a regression for the GFDL-CM3 \citep{Griffies2011a} for the grid cell containing New York City for both summertime monthly mean temperature and precipitation rate (RCP 8.5). For local precipitation patterns, unforced variability is greater, and there is a weaker correlation with global mean temperature. 

We use a single realization from each CMIP5 model, but note that, for models where multiple realizations are available, fitting the output from additional model realizations could more tightly constrain the separation into both forced changes and unforced variability and could allow for alternative approaches in which $\hat{k}$ is not constant with temperature. Other approaches might also include additional covariates, such as aerosol emissions which can modify the patterns \citep[e.g.,][]{Frieler2012a}. Maps of each model's annually-derived temperature and precipitation patterns for the CONUS are shown in Figs. A3 and A4. For temperature, most models have similar patterns. Larger inter-model differences for precipitation have been suggested to originate from large background variations in precipitation masking the forced signal \citep{Tebaldi2011a,Hawkins2011a}. 

\subsection{Probability weighting}

\subsubsection{Equal-weighted CMIP5 ensemble}

As a baseline for comparing the SMME and MCPR projections, we employ an equal-weighted CMIP5 ensemble. In interpreting this ensemble, we follow the approach of the Intergovernmental Panel on Climate Change (IPCC). In particular, we note that, while in IPCC terminology the phrases \emph{very likely} and \emph{likely} bracket the 5 to 95th percentile and the 17 to 83rd percentile outcomes, respectively, the IPCC's Fifth Assessment Report (AR5) uses the 5 to 95th percentile range of long-term temperature change as projected by CMIP5 to bound the \emph{likely} outcome \citep[see][section 12.4.1.1]{Collins2013a}. The underlying judgement that the CMIP5 archive does not adequately represent the tails of projected future temperature change is based upon the observation that the \emph{likely} (17th/83rd) range of the transient climate response (TCR) \citep{Cubasch2001a}, inferred from multiple lines of evidence, corresponds to the 5 to 95th percentile range of the TCR from the CMIP5 models \citep{Collins2013a}, as well as more general informal expert assessment of confidence in GCM projections. Consequently, we compare the 5th to 95th percentile of temperature projections from the equal-weighted CMIP5 ensemble to the 17th to 83rd percentile range of the probability distributions from the SMME and MCPR methods. For precipitation projections, we do not make such an adjustment.

\subsubsection{Surrogate/Model Mixed Ensemble (SMME) method}

In the Surrogate/Model Mixed Ensemble (SMME) method, we divide the unit interval $[0,1]$ into ten bins, with a somewhat higher density of bins at the tails of the interval to ensure sampling. Specifically, the bins are centered at the 4th, 10th, 16th, 30th, 50th, 70th, 84th, 90th, 94th, and 99th percentiles. The quantiles of global mean temperature change corresponding to the bounds and center of each bin are taken from the MAGICC6 output. CMIP5 model output is categorized into bins based on the projected change in global mean temperature from 1981--2010 to 2080--2099.

In bins, primarily at the tail of the distribution, not represented by at least two CMIP5 models, we generate model surrogates sufficient to bring the number of models plus surrogates to two. To generate a model surrogate, we start from the global mean temperature time series of the MAGICC6 projection that corresponds to the central quantile of the bin. If there is no CMIP5 output in the bin, we pick two models with global mean temperature projections close to the bin, such that one model pattern reflects a net increase in CONUS precipitation with temperature and one reflects a net decrease (or lesser increase) in CONUS precipitation with temperature. If there is a single CMIP5 model in the bin, we pick a single model with a precipitation pattern either identical or complementing the one in the bin. We then use the patterns from the selected models to scale the global mean temperature projection and add the residuals from the same models, generating a surrogate model that includes both forced change and unforced variability. Tables A3-A6 list the models used to generate each pattern as well as their respective global mean temperature bin assignment.

In the final probability distribution, the models and surrogates in a bin are weighted equally such that the total weight of the bin corresponds to the target distribution for 2080--2099 temperature. For example, if there are four models in the bin centered at the 30th percentile and stretching from the 20th to the 40th percentiles, each will be assigned a probability of $20\%/4 = 5\%$. Thus the projected distribution for global mean temperature approximates the target (Figs. \ref{fig:global_tas_rcp85}, A5-A7).

\subsubsection{Monte Carlo Pattern/Residual (MCPR) method}

In the Monte Carlo Pattern/Residual (MCPR) method, we use the CMIP5 output as a source of patterns and residuals but do not directly retain any model output. Instead, we divide the unit interval $[0,1]$ into 100 equal bins and take the quantile of MAGICC6 global mean temperature projections corresponding to the center of each bin (i.e., the 0.5th, 1.5th, 2.5th, etc., percentiles). We generate a pool of candidate patterns by replicating the list of patterns a sufficient number of times to meet or exceed the number needed, then sample without replacement from the pool to assign a pattern to each bin. We sample without replacement from an identical pool to assign a residual time series to each bin. We then use the global mean temperature projection, the pattern, and the residual time series to generate a projection for each bin. Each projection is of equal probability, but patterns and/or residuals could be alternatively weighted (e.g., historical performance, pattern accuracy in reproducing GCM results). The identical pairs of CMIP5 patterns and residuals are assigned to each bin to project both temperature and precipitation.

\section{Results}
\label{results}

\subsection{Temperature projections}

As expected, the equal-weighted CMIP5 ensemble fails to produce the upper tail of MAGICC6 global temperature distribution; at the 95th percentile, the CMIP5  projection for RCP 8.5 in 2080--2099 is $\sim$2$^\circ$C cooler than the MCPR and SMME methods. However, for the lower tail and center of the cumulative distribution function (CDF), the three methods are generally within 1$^\circ$C of one another (Fig. \ref{fig:tas_prCDF}B).

In ocean-atmosphere coupled global climate models, the land commonly warms faster than the ocean \citep[e.g.,][]{Manabe1990}. As such, the CONUS temperature increases are projected to continue to be greater than the global mean. For the CONUS, the upper tail from the probabilistic methods is also not well captured by the CMIP5 ensemble; the 95th percentile from the CMIP5 is $\sim$1$^\circ$C less than that of the MCPR and SMME methods (RCP 8.5, 2080-2099) (Fig. \ref{fig:tas_prCDF}D). However, there is agreement between methods at the 95th percentile under the lowest emissions pathway (RCP 2.6, 2080--2099): 2.6$^\circ$C (CMIP5), 2.6$^\circ$C (MCPR), 2.7$^\circ$C (SMME) (Table \ref{tab:tasannual2080-2099}). Additionally, all methods generally agree with the \emph{likely} (17th--83rd for SMME and MCPR; 5th--95th for CMIP5) range of 2080--2099 projected CONUS temperatures (RCP 8.5): 3.3--6.9$^\circ$C (CMIP5), 3.4--6.9$^\circ$C (SMME), and 3.5--6.5$^\circ$C (MCPR) (Table \ref{tab:tasannual2080-2099}). For CONUS sub-regions (defined in Fig. A8), the \emph{likely} range from the probability distributions (17th/83rd) is generally within a half-degree of the CMIP5 ensemble (5th/95th) (Table \ref{tab:tasannual2080-2099}).

By the end of the century under RCP 8.5, all methods project very similar \emph{likely} ranges (17th--83rd for SMME and MCPR; 5th--95th for CMIP5) of June-July-August (JJA) CONUS temperature increase: 3.8--7.4$^\circ$C (CMIP5), 3.9--7.4$^\circ$C (SMME), and 3.9--7.3$^\circ$C (MCPR), with a 5\% chance that average JJA temperatures could rise by as much as 8.2 to 9.3$^\circ$C (SMME and MCPR) (Table A7). Overall, late century 5th and 50th percentile geographic patterns of warming are comparable between methods, generally $<$ 1$^\circ$C difference in most areas (both December-January-February [DJF] and JJA) (Figs. \ref{fig:wintertasmap} and \ref{fig:summertasmap}, respectively). The greatest 5th and 50th percentile JJA warming occurs over the Upper Great Plains, the Upper Midwest and areas over the mountain states in theunforced variability western U.S. These areas, in addition to Alaska and New England, also warm the most during DJF by the end of century and are relatively consistent between the three ensembles at the 5th and 50th percentiles. At the 95th percentile, JJA temperature projections of the SMME and MCPR methods are similar, with much of the CONUS and Alaska experiencing at least a 9$^\circ$C rise in temperature by the end of the century. However, there is more disagreement for DJF; 95th percentile DJF temperature increases from the MCPR method are roughly 1 to 4$^\circ$C warmer than the SMME method over the Great Plains and the Upper Midwest.

To compare the influence of the different methods on projections of temperature extremes, we estimate the number of  ``extremely warm'' days where the maximum temperature  above 35$^\circ$C  and the number of ``extremely cold'' days where the minimum temperature is below 0$^\circ$C. Taking a population-weighted average of historical county-level daily maximum temperatures, we estimate that the average American experiences nearly 15 days each year when the maximum temperature is \textgreater 35$^\circ$C and 74 days when the minimum temperature is \textless 0$^\circ$C (1981--2010). By 2080--2099 under RCP 8.5, the CMIP5, MCPR, and SMME methods all project that the number of extremely warm days will \emph{likely} (17th--83rd for SMME and MCPR; 5th--95th for CMIP5) more than triple -- a rate faster than that of annual temperatures (Table A8), while all methods agree that the number of extremely cold days will \emph{likely} be reduced by one-half. Spatially, very little differences exist between methods in the expected (i.e., weighted ensemble average) number of projected extremely warm and cold temperature days (Figs. A9 and A10, respectively). The MCPR and SMME methods suggest there is a 5\% chance that the current number of extremely warm temperature days could increase almost eightfold (Table A8), while extremely cold temperature days could decline $\sim$75\% (Table A9). By comparison, the hottest CMIP5 model projects roughly a sevenfold increase in extremely warm days and a $\sim$64\% decrease in extremely cold days (Fig. A11).

\subsection{Precipitation projections}

For all methods considering precipitation, we define the \emph{likely} range as the 17th--83rd and the \emph{very likely} range as the 5th--95th. By the end of the 21st century, CONUS annual precipitation will \emph{likely} (67\% probability, MCPR, SMME and CMIP5) increase (Table \ref{tab:prcpannual2080-2099}). Additionally, all methods project that the Northeast, Midwest, and Upper Great Plains are \emph{likely} to experience more winter precipitation around the same time (RCP 8.5) (Table A10). We also find that wetter springs are \emph{very likely} (90\% probability, MCPR, SMME and CMIP5) in the Northeast, Midwest, and Upper Great Plains, and \emph{likely} in the Northwest and Southeast (MCPR, SMME and CMIP5; Table A11). An increase in fall precipitation is \emph{likely} in the Northeast, Midwest, Upper Great Plains, and Southeast. In general, many of the CMIP5 models project mid- and high latitude precipitation increases, with changes becoming more pronounced as temperature increases \citep[see][Figs. 12.10 and 12.22]{Collins2013a}. The MCPR, SMME and CMIP5 project that the Southwest is \emph{likely} to experience drier springs, while drier summers are \emph{likely} in the Great Plains and the Northwest (Table A12). The CMIP5 projects slightly drier average spring conditions in the Southwest than the probabilistic ensembles (Fig. A12 and Figs. \ref{fig:tas_prCDF}E,F), but for other regions and time periods the median precipitation projections from the SMME are slightly drier than the CMIP5 and MCPR.

\subsection{Sources of projection uncertainty}

For decision making purposes, it is useful to examine future climate change projection uncertainty, which can be decomposed into: 1) forced, 2) unforced, and 3) scenario (i.e., emissions) uncertainty -- each of which can evolve with time and location \citep[e.g.,][]{Hawkins2009a}. Similar to \citet{Hawkins2009a}, we estimate the evolution of the fractional contribution of all three uncertainty components over the 21st century for global and local scales. However, while \citet{Hawkins2009a} assumes unforced variability is time invariant (estimated as the residual from a 4th-order polynomial fit to modeled regional and global mean temperatures), we instead use the time series of unforced variability calculated from pattern scaling. (Methods for all component uncertainty calculations are described in Appendix A.)

Figure \ref{fig:variance} shows the relative importance of each of the three uncertainty components for annual temperature globally and in four illustrative locations (Los Angeles, California; New Orleans, Louisiana; Portland, Maine; Seattle, Washington). The year 2000 is chosen as the reference point. Globally, unforced variability dominates in the near-term, but falls to less than half of total variance around 2020.  Scenario uncertainty becomes larger than uncertainty in the forced response around 2060 (Fig. \ref{fig:variance}A). For all four locations, up until the middle of the 21st century, projection uncertainty from unforced variability dominates. Only in the 2050s--2060s, as the variance associated with uncertainty in the forced change and in the scenario increases, does the variance from unforced variability fall to less than half the total. Consistent with regional breakdowns from \citet{Hawkins2009a}, there is very little projection uncertainty associated with emissions scenarios until the 2040s.

\subsection{Projection uncertainty due to unforced variability}

Even at the global scale, the forced climate change signal can sometimes be masked by unforced variability. In most multi-model studies, a single realization of each GCM is used for the primary purpose of identifying forced trends. By contrast, several runs of the same model initialized from different initial states of the atmosphere can yield multiple estimates of weather for any given year. If external forcing is constant, differences between simulations are solely attributed to internal variability. While computationally expensive, these ensembles can estimate near-term projection uncertainty due to year-to-year fluctuations in weather \citep[e.g.,][]{Kay2014a, Deser2014a}.

For example, \citet{Kay2014a} construct a 30-member ensemble with the CESM1-CAM5 model \citep{Meehl2013a,Hurrell2013a}. Each member simulation uses slightly different atmospheric initial conditions, while the external anthropogenic forcing remains constant (RCP 8.5). The authors calculate 10 and 20-year global temperature trends starting from every year from 1990 to 2009 and from 2030 to 2049 and then construct histograms of the trends. The spread of each distribution is an estimate of projection uncertainty due to unforced variability (Fig. \ref{fig:kayCMIP5}, red histogram). 

For a particular prescribed forcing, \citet{Kay2014a} note that the temperature projection spread of the 30-member CESM1-CAM5 ensemble aligns closely with that of the spread of an ensemble of CMIP5 models (both their own forced and unforced components). As an extension, we further assess whether superimposing just the unforced temperature projection components from an ensemble of CMIP5 models with the CESM1-CAM5 forced component (RCP 8.5) produces a similar range of unforced variability. To do this, we add the unforced variability component of global temperature from each model (and model surrogate in the case of SMME and MCPR methods) to the CESM1-CAM5 forced component (Fig. \ref{fig:kayCMIP5}). The resulting distributions of trends from both methods closely align with those from \citet{Kay2014a} (red). Our approach may be beneficial for estimating projection uncertainty from unforced variability when computational resources are not available to facilitate additional ensemble simulations. While these results are global, regional climates are generally more impacted by unforced variability \citep{Kay2014a}. Therefore, further investigation should consider how well records of unforced variability from the CMIP5 reproduce the spread of local trends from multi-member initial condition ensembles.
 
\section{Discussion} \label{discussion}

Both the SMME and MCPR methods generate joint probability distributions of temperature and precipitation that originate from a prescribed PDF of global mean temperature. These are joint PDFs because for each realization, we source the temperature and precipitation forced and unforced components from the same GCM. In contrast to the equal-weighted CMIP5 ensemble, which also generates joint estimates, the SMME and MCPR projections are consistent with probabilistic global mean temperature projections. The particular global mean temperature projections used considers a distribution of model parameters consistent with both historical observations and the IPCC's consensus on equilibrium climate sensitivity, and thus allows sampling of low-probability outcomes that are outside the range of GCM ensembles. Accordingly, the results of the SMME and MCPR methods are well suited for use in probabilistic risk analyses, and are particularly ripe for integration with sector-specific impact models and damage functions \citep[e.g.,][]{Deschnes2011a,Auffhammer2011a,Houser2015a}, including those jointly dependent on temperature and precipitation \citep[e.g.,][]{Schlenker2009a}. Probabilisitic projections facilitate impact estimates that incorporate physical climate projection uncertainty, which may be especially useful for decision making under uncertainty.  Furthermore, the decomposition of projection variance illustrates the importance of including unforced variability in estimates of future climate change. Applying impact functions based solely on forced changes would omit the primary driver to annual temperature uncertainty through the middle of the 21st century (Fig. \ref{fig:variance}).

The SMME and MCPR approaches span the range of possibilities regarding the correlation between GCM projections of forced changes and GCM projections of unforced change. The SMME approach assumes that these are perfectly correlated -- the projected forced pattern from a given model is always used with the unforced residuals from the same model. The MCPR approach, by contrast, assumes that these are fully decoupled. Nevertheless, despite the differences in the approaches between the SMME and MCPR methods, there are few instances where the distribution of 20-year average local temperature and precipitation projections substantially deviate from one another.

Relative to temperature, regional precipitation estimates exhibit a wider range of outcomes in both direction and magnitude of changes. This is likely due to disagreement in the response to anthropogenic forcing over the U.S. across GCMs (Fig. A4). While future model development efforts should address these disagreements, current approaches that may narrow the range of outcomes include alternative model weighting schemes, such as model weights based in part on historical precipitation performance rather than projected global mean temperature. Another example of projection disagreement is the late 21st century median CONUS temperature anomaly (RCP 8.5), in which the MCPR projection is $\about$0.5$^\circ$C cooler than CMIP5 and SMME projections (Figs. A13). This difference may be due to the MCPR method selecting a greater number of models that have a cooler average forced temperature pattern over the CONUS. 

Both the SMME and (especially) MCPR methods rely upon pattern scaling, the limitations of which have been extensively summarized by \citet{Tebaldi2014a}. These limitations should be kept in mind when interpreting these results, particularly that forced patterns represent long-term averages of climate parameters and may omit non-linear effects such as climate feedbacks that could alter rates of warming. Likewise, pattern scaling is intended for scenarios with continuously increasing forcing. For strong mitigation scenarios where forcing can be increasing and then decreasing (e.g., RCP 2.6), separate patterns for each pathway may be more appropriate.

It is important to stress that these results are conditional upon one particular PDF of global mean temperature change. These same methods from probabalizing the CMIP5 projections can, however, be employed with any PDF of global temperature change. Moreover, in the presence of deep uncertainty, it might be appropriate to apply more than one probability distribution using methods that rely on multiple priors \citep[e.g.,][]{Heal2014a}. Both the SMME and MCPR methods could be implemented in such a framework.

Some climate risks may be less amenable to probabilistic analysis based on PDFs like those produced here, and may instead require scenario-based, ``possibilistic'' analysis \citep[e.g.,][]{Whiteman2013a}. These include risks arising from feedbacks that might amplify global mean temperature increase that are not captured in the SCM, such as omitted carbon cycle feedbacks that include the release of methane from permafrost of hydrates \citep{Archer2007a}. These also include risks arising from factors affecting local projections that are poorly captured in GCMs, such as mid-latitude extremes that may be influenced by the failure to properly pace Arctic sea ice loss \citep{Francis2012a}.

\section{Conclusion} \label{conclusion}

While projections from GCM ensembles like those produced by CMIP5 characterize \emph{likely} (17th--83rd percentile) range of temperature  and precipitation change, they undersample extreme behavior, which may be critical for effective risk management. In this study, we present two alternative approaches for generating time series of joint probabilistic projections of temperature and precipitation that include tail risk. Projections from both probabilisitic methods and an equal-weighted GCM ensemble are available online and are summarized in the text and appendices for both multiple lead times and U.S. sub-regions.

The CMIP5 models substantially underestimate the 95th percentile projections from the probabilisitic methods. We find that by the end of the 21st century, there is a 5\% chance that annual CONUS temperature change could be as high as $\sim$8$^\circ$C over 1981--2010 levels -- roughly 1$^\circ$C warmer than the hottest CMIP5 model projection (RCP 8.5). We also find that there is a 5\% chance that the average American could experience nearly 4 months out of the year when daily maximum temperature is 30$^\circ$C or warmer. However, strong CO$_2$ emissions mitigation can greatly reduce these risks. Under RCP 2.6, we project that increases in CONUS temperature will \emph{very likely} (90\% probability) remain at or under 2.7$^\circ$C by the end of the century and the number of extremely warm days experienced by the average American could coincidently remain below $\sim$40 days per year.

Decomposing GCM output into forced and unforced components of climate change via pattern scaling can provide records useful for uncertainty quantification. We find that uncertainties associated with local temperature projections through 2050 are almost entirely due to unforced variability, with a small fraction arising from uncertainty in the forced component of climate change. By the end of the 21st century, uncertainty associated with CO$_{2}$ emissions dominates both at global and local scales. 

%
%

\acknowledgments
We thank M. Oppenheimer for helpful discussion. DMR and REK were supported by the Risky Business Project and by the Global Climate Prospectus through the University of Chicago 1896 Fund. We acknowledge the World Climate Research Programme's Working Group on Coupled Modeling, which is responsible for CMIP, and we thank the climate modeling groups (listed in Table A2 of this paper) for producing and making available their model output. For CMIP the U.S. Department of Energy's Program for Climate Model Diagnosis and Intercomparison provides coordinating support and led development of software infrastructure in partnership with the Global Organization for Earth System Science Portals.


 \bibliographystyle{ametsoc2014}
 \bibliography{TP}


\clearpage
\newpage

\begingroup
\setlength{\tabcolsep}{3pt}
\renewcommand*{\arraystretch}{1}

\setcounter{table}{0}
\renewcommand{\thetable}{\arabic{table}}

\clearpage \begin{table*}[p]
\centering
\caption{Projected regional annual temperature change ($^\circ$C) under all RCPs for 2080-2099}\label{tab:tasannual2080-2099}
\begin{tabular}{l|rrr|rrr|rrr|rrr}
 Annual & \multicolumn{3}{c|}{RCP 8.5} & \multicolumn{3}{c|}{RCP 6.0} & \multicolumn{3}{c|}{RCP 4.5} & \multicolumn{3}{c}{RCP 2.6} \\ 
2080-2099& 50 & 17--83 & 5--95 & 50 & 17--83 & 5--95 & 50 &17--83 & 5--95 & 50 & 17--83 & 5--95 \\ 
\hline \multicolumn{13}{l}{CONUS (1981--2010 normal: 11.9$^\circ$C)} \\
CMIP5&5.1&3.6--5.9&3.3--6.9&3.2&2.4--3.7&1.6--4.5&2.3&1.6--3.3&1.5--3.8&1.5&0.7--2.0&0.7--2.6\\
SMME&5.2&3.4--6.9&3.2--8.0&3.2&2.3--4.3&1.4--5.3&2.4&1.6--3.8&1.5--4.8&1.3&0.6--2.0&0.5--2.7\\
MCPR&4.5&3.5--6.5&3.1--8.3&3.1&2.2--4.5&1.9--5.7&2.4&1.7--3.5&1.3--4.9&1.3&0.8--2.0&0.6--2.6\\
\hline \multicolumn{13}{l}{Northeast (1981--2010 normal: 8.8$^\circ$C)} \\
CMIP5&5.2&3.8--6.3&3.6--7.0&3.2&2.4--4.0&1.6--4.8&2.5&1.6--3.6&1.4--4.1&1.7&0.7--2.1&0.7--2.7\\
SMME&5.2&3.7--7.1&3.1--8.8&3.2&2.1--4.6&1.5--5.5&2.6&1.6--4.0&1.4--4.6&1.4&0.6--2.1&0.3--3.1\\
MCPR&4.8&3.7--7.0&3.2--8.8&3.2&2.2--4.7&1.7--5.7&2.5&1.8--3.6&1.4--5.3&1.4&0.8--2.2&0.4--2.8\\
\hline \multicolumn{13}{l}{Southeast (1981--2010 normal: 16.6$^\circ$C)} \\
CMIP5&4.4&3.5--5.2&2.7--5.9&2.7&2.2--3.2&1.4--3.7&2.1&1.4--2.6&1.3--3.2&1.3&0.6--1.6&0.5--1.8\\
SMME&4.4&3.0--5.9&2.6--7.3&2.7&2.2--3.7&1.3--4.6&2.2&1.4--3.2&1.3--3.8&1.1&0.5--1.6&0.4--2.3\\
MCPR&3.9&3.0--6.0&2.8--7.3&2.7&1.9--3.8&1.5--4.5&2.0&1.4--3.0&1.1--4.1&1.0&0.5--1.6&0.3--2.2\\
\hline \multicolumn{13}{l}{South Central (1981--2010 normal: 17.8$^\circ$C)} \\
CMIP5&4.8&3.9--5.7&3.3--6.2&2.8&2.5--3.6&1.4--3.9&2.3&1.4--3.0&1.2--3.5&1.3&0.9--1.8&0.4--2.1\\
SMME&4.9&3.3--6.2&3.1--7.6&2.8&2.5--3.9&1.3--4.8&2.3&1.8--3.5&1.2--4.4&1.2&0.4--1.8&0.4--2.5\\
MCPR&4.5&3.4--6.5&3.1--8.0&3.0&2.1--4.1&1.8--5.1&2.2&1.6--3.6&1.1--4.5&1.3&0.7--1.9&0.3--2.5\\
\hline \multicolumn{13}{l}{Upper Great Plains (1981--2010 normal: 8.9$^\circ$C)} \\
CMIP5&5.4&3.7--6.5&3.1--7.3&3.5&2.6--4.2&1.6--4.8&2.6&1.6--3.8&1.5--4.4&1.6&0.9--2.4&0.7--2.8\\
SMME&5.5&3.5--7.3&3.2--8.2&3.5&2.1--4.7&1.5--6.0&2.8&1.6--4.4&1.6--5.6&1.5&0.7--2.4&0.6--2.8\\
MCPR&4.9&3.7--7.2&3.3--9.0&3.4&2.4--4.8&1.8--6.0&2.5&1.8--3.7&1.5--5.5&1.5&0.9--2.4&0.6--3.0\\
\hline \multicolumn{13}{l}{Midwest (1981--2010 normal: 9.0$^\circ$C)} \\
CMIP5&5.4&3.8--6.5&3.5--7.9&3.6&2.6--4.2&1.8--5.0&2.7&1.7--3.7&1.5--4.5&1.6&0.8--2.2&0.7--3.1\\
SMME&5.7&3.5--7.9&3.4--8.7&3.6&2.2--4.7&1.6--5.9&2.7&1.6--4.5&1.5--5.8&1.6&0.7--2.2&0.6--3.1\\
MCPR&4.9&3.8--7.3&3.4--9.3&3.4&2.4--4.9&1.9--6.1&2.6&1.9--3.9&1.5--5.5&1.4&0.9--2.3&0.7--3.2\\
\hline \multicolumn{13}{l}{Northwest (1981--2010 normal: 9.5$^\circ$C)} \\
CMIP5&4.3&3.2--5.8&3.1--6.4&2.9&1.9--3.7&1.5--4.3&2.2&1.3--3.2&1.0--3.6&1.6&1.1--2.1&0.7--2.7\\
SMME&4.3&3.1--6.4&2.5--7.1&2.9&1.6--4.3&1.5--5.1&2.3&1.4--3.6&1.3--4.7&1.3&1.0--2.1&0.7--2.9\\
MCPR&4.3&3.3--6.3&2.9--7.7&2.9&2.0--4.1&1.6--5.1&2.2&1.4--3.3&1.1--5.0&1.3&0.8--2.1&0.6--2.9\\
\hline \multicolumn{13}{l}{California (1981--2010 normal: 15.5$^\circ$C)} \\
CMIP5&4.1&3.4--5.1&2.9--6.0&2.6&2.0--3.4&1.3--3.8&2.1&1.4--2.7&1.3--3.1&1.3&1.0--1.8&0.7--2.3\\
SMME&4.2&3.0--6.0&2.9--6.8&2.6&1.9--3.7&1.2--4.6&2.2&1.4--3.1&1.4--4.0&1.2&0.8--1.8&0.3--2.3\\
MCPR&4.1&3.1--5.7&2.6--7.3&2.6&1.9--3.9&1.7--4.8&2.1&1.4--3.1&1.1--4.3&1.2&0.8--1.8&0.6--2.4\\
\hline \multicolumn{13}{l}{Southwest (1981--2010 normal: 14.2$^\circ$C)} \\
CMIP5&4.7&4.0--6.0&3.6--6.9&2.9&2.5--3.6&1.5--4.4&2.4&1.6--3.2&1.5--3.7&1.3&0.9--1.8&0.7--2.5\\
SMME&4.7&3.9--6.9&3.5--8.4&2.9&2.3--4.1&1.4--5.0&2.5&1.7--3.7&1.5--4.6&1.2&0.8--1.8&0.5--2.5\\
MCPR&4.7&3.5--6.6&3.1--8.5&3.1&2.2--4.2&1.9--5.4&2.3&1.6--3.5&1.2--4.6&1.4&0.8--2.0&0.6--2.4\\
\hline \multicolumn{13}{l}{Rocky Mountain States (1981--2010 normal: 8.1$^\circ$C)} \\
CMIP5&5.3&3.8--6.2&3.4--7.5&3.4&2.3--3.9&1.6--4.9&2.5&1.6--3.5&1.4--4.2&1.6&0.9--2.2&0.7--3.0\\
SMME&5.3&3.5--7.5&3.3--8.4&3.4&2.2--4.5&1.5--5.5&2.6&1.6--4.2&1.5--5.4&1.3&0.8--2.2&0.6--3.0\\
MCPR&4.7&3.8--7.0&3.1--8.9&3.2&2.3--4.7&2.0--5.8&2.4&1.8--3.8&1.4--5.3&1.4&0.9--2.2&0.7--2.9\\
\hline \multicolumn{13}{l}{Alaska (1981--2010 normal: -1.7$^\circ$C)} \\
CMIP5&5.8&4.2--8.2&3.7--9.0&3.8&3.1--4.9&2.4--5.8&3.0&1.9--4.3&1.9--5.2&2.3&1.4--3.1&0.4--3.5\\
SMME&6.3&4.2--9.1&3.7--10.6&3.8&2.7--5.7&2.4--7.1&3.2&2.0--5.2&1.9--6.6&2.0&0.8--3.1&0.7--3.6\\
MCPR&6.2&4.5--8.8&4.0--12.5&4.0&2.7--5.8&2.3--7.4&3.2&2.1--4.7&1.5--6.9&1.8&1.0--2.9&0.5--4.2\\
\hline \multicolumn{13}{l}{Hawaii (1981--2010 normal: 23.6$^\circ$C)} \\
CMIP5&2.8&2.3--4.1&2.1--4.5&1.6&1.2--2.7&1.2--3.0&1.3&1.0--1.9&0.9--2.6&0.9&0.5--1.3&0.5--1.9\\
SMME&2.9&2.1--4.3&1.8--5.2&1.6&1.2--2.8&1.2--3.6&1.4&1.1--2.3&0.9--2.8&0.8&0.5--1.3&0.4--1.9\\
MCPR&2.9&2.3--4.3&2.1--5.2&1.8&1.3--2.6&1.1--3.5&1.5&1.0--2.3&0.7--3.0&0.8&0.5--1.4&0.3--2.0\\
\hline
\end{tabular}
\end{table*}

\clearpage \begin{table*}[p]
\centering
\caption{Projected regional annual precipitation change (\%) under all RCPs for 2080-2099}\label{tab:prcpannual2080-2099}
 \begin{tabular}{l|rrr|rrr|rrr|rrr}
Annual & \multicolumn{3}{c|}{RCP 8.5} & \multicolumn{3}{c|}{RCP 6.0} & \multicolumn{3}{c|}{RCP 4.5} & \multicolumn{3}{c}{RCP 2.6} \\
2080-2099&50 & 17--83 & 5--95 & 50 & 17--83 & 5--95 & 50 &17--83 & 5--95 & 50 & 17--83 & 5--95 \\
\hline \multicolumn{13}{l}{CONUS (1981--2010 normal: 750.4mm)} \\
CMIP5&6.8&-0.8--12.5&-5.4--15.0&3.0&-0.1--7.2&-2.1--16.8&3.8&0.3--7.8&-2.6--10.7&3.1&-0.9--8.5&-3.1--11.0\\
SMME&2.0&-0.3--10.7&-5.4--13.5&2.4&-0.3--7.2&-0.9--14.2&2.7&-0.4--7.1&-1.2--9.1&2.6&-0.3--7.9&-2.5--11.0\\
MCPR&7.8&0.3--13.7&-4.6--21.7&4.8&-0.4--9.3&-1.8--17.5&5.3&0.5--9.2&-2.0--13.8&3.0&-1.2--7.4&-4.3--11.4\\
\hline \multicolumn{13}{l}{Northeast (1981--2010 normal: 1103.7mm)} \\
CMIP5&11.5&6.2--16.4&4.1--18.6&9.4&3.4--13.1&0.2--18.1&6.9&2.6--11.7&-1.0--14.3&3.5&0.4--10.2&-1.1--11.7\\
SMME&9.6&0.2--15.2&-0.2--18.5&6.8&0.0--12.5&-0.3--17.1&5.6&-0.1--10.2&-0.3--14.3&2.8&-0.4--8.3&-1.1--11.2\\
MCPR&12.4&7.2--19.8&3.2--25.1&9.7&5.0--16.4&0.7--21.9&7.8&4.3--14.2&1.0--18.2&3.2&0.1--8.6&-2.3--15.2\\
\hline \multicolumn{13}{l}{Southeast (1981--2010 normal: 1303.9mm)} \\
CMIP5&4.9&-3.5--15.3&-17.3--23.6&3.0&-0.1--12.4&-3.5--17.0&4.1&-0.2--10.6&-7.6--16.5&3.0&-0.3--10.1&-5.9--15.3\\
SMME&1.2&-2.0--13.8&-11.5--21.9&1.3&-0.6--12.4&-3.5--17.0&3.4&-0.4--9.2&-1.7--13.5&2.5&-0.3--7.1&-4.7--15.3\\
MCPR&7.4&-1.5--21.6&-5.2--28.6&3.7&-0.1--15.5&-3.4--20.8&6.7&0.9--12.2&-2.5--17.6&3.9&-1.0--8.8&-5.9--14.9\\
\hline \multicolumn{13}{l}{South Central (1981--2010 normal: 918.7mm)} \\
CMIP5&-1.7&-14.3--6.0&-19.8--11.7&-2.2&-9.8--3.9&-12.8--13.3&-1.4&-4.4--5.5&-8.4--12.5&3.0&-3.8--7.1&-12.2--8.3\\
SMME&-0.3&-7.5--5.1&-19.8--10.5&-0.6&-9.4--3.2&-12.8--5.1&-0.8&-3.4--4.9&-8.4--12.5&0.8&-2.8--6.9&-6.9--8.3\\
MCPR&-0.9&-11.2--8.1&-21.0--12.6&-2.4&-9.4--4.6&-14.8--9.9&-0.1&-7.5--7.2&-13.5--10.5&1.8&-6.1--7.7&-11.2--13.1\\
\hline \multicolumn{13}{l}{Upper Great Plains (1981--2010 normal: 588.6mm)} \\
CMIP5&4.0&-1.5--13.7&-5.4--20.0&1.0&-3.3--6.1&-6.4--17.5&3.5&-1.8--7.2&-6.4--12.2&4.8&-1.6--7.2&-4.5--10.3\\
SMME&1.8&-0.1--10.1&-5.1--20.0&0.4&-0.8--6.1&-5.8--17.5&2.8&-0.4--7.0&-6.4--12.2&2.0&-0.5--6.7&-4.5--10.3\\
MCPR&5.7&-0.1--14.0&-9.3--26.1&3.1&-3.2--10.0&-5.1--18.5&4.8&-2.1--11.3&-6.1--18.4&1.4&-3.7--7.9&-9.7--13.8\\
\hline \multicolumn{13}{l}{Midwest (1981--2010 normal: 923.4mm)} \\
CMIP5&10.1&4.4--14.3&-2.4--21.1&5.4&-0.5--11.3&-1.4--16.3&5.4&0.7--9.2&-4.0--14.2&3.9&-0.3--7.4&-2.4--9.3\\
SMME&6.3&0.3--14.0&-0.6--20.2&2.3&-0.1--8.8&-0.8--16.3&4.8&-0.1--8.7&-4.0--10.8&3.4&-0.2--7.3&-2.4--9.3\\
MCPR&8.5&3.0--16.6&-1.3--23.4&5.1&-0.8--11.7&-3.1--19.1&6.2&0.8--12.2&-1.1--16.9&2.2&-1.8--5.2&-4.8--10.9\\
\hline \multicolumn{13}{l}{Northwest (1981--2010 normal: 767.8mm)} \\
CMIP5&5.1&0.5--13.6&-7.8--18.2&7.3&2.3--10.6&-1.0--17.0&5.2&-1.4--10.0&-3.2--15.1&3.8&-2.4--5.4&-3.8--12.6\\
SMME&2.1&-0.8--11.6&-7.8--17.2&5.6&-0.7--9.8&-1.0--11.5&2.6&-1.1--9.3&-3.2--15.1&0.6&-2.4--5.1&-3.8--9.2\\
MCPR&7.5&-1.7--15.0&-10.3--22.6&6.7&-0.9--13.4&-4.5--22.9&6.1&-0.8--13.7&-5.1--18.9&1.9&-3.9--8.6&-8.0--14.5\\
\hline \multicolumn{13}{l}{California (1981--2010 normal: 530.4mm)} \\
CMIP5&3.0&-11.2--14.3&-26.5--32.9&3.7&-8.2--16.2&-19.3--21.7&0.5&-8.7--15.2&-15.3--19.7&1.7&-7.5--11.6&-14.6--15.7\\
SMME&-1.2&-10.9--13.5&-26.5--28.5&0.0&-8.2--13.4&-15.5--21.4&-0.9&-7.6--10.1&-15.3--19.7&-0.7&-3.1--9.5&-14.6--15.7\\
MCPR&6.0&-8.0--23.3&-20.3--34.6&5.0&-13.2--17.9&-22.8--22.3&2.5&-11.6--13.4&-22.2--23.0&4.1&-8.7--16.0&-16.9--23.8\\
\hline \multicolumn{13}{l}{Southwest (1981--2010 normal: 312.1mm)} \\
CMIP5&-1.4&-8.2--11.6&-24.5--13.1&-3.2&-13.5--7.6&-19.3--10.2&-0.1&-7.6--8.4&-14.4--10.1&-0.3&-9.2--7.7&-11.0--16.4\\
SMME&-0.4&-7.6--7.5&-22.2--12.8&-0.4&-11.5--5.6&-15.4--10.2&-0.3&-7.4--7.4&-14.4--10.1&-0.3&-9.2--6.4&-9.3--16.4\\
MCPR&-0.2&-16.4--12.4&-31.1--16.5&-3.5&-14.6--7.7&-21.8--16.3&-0.9&-12.9--9.3&-28.9--16.5&3.5&-9.0--11.3&-21.6--21.7\\
\hline \multicolumn{13}{l}{Rocky Mountain States (1981--2010 normal: 330.6mm)} \\
CMIP5&11.5&3.7--21.9&-6.4--22.5&9.1&2.8--17.9&-2.0--25.6&8.9&-0.6--12.4&-5.8--18.6&7.7&2.3--12.4&-5.3--15.4\\
SMME&8.7&0.1--17.5&-3.1--22.5&4.3&0.1--14.7&-0.7--22.2&4.2&-0.6--10.1&-5.7--18.6&2.7&-0.3--11.3&-5.3--15.4\\
MCPR&15.1&1.5--26.5&-1.9--35.3&11.2&1.9--22.9&-3.5--29.4&9.1&-0.1--17.8&-6.1--24.2&6.2&-2.6--13.9&-7.9--19.6\\
\hline \multicolumn{13}{l}{Alaska (1981--2010 normal: 541.2mm)} \\
CMIP5&31.5&25.5--38.6&18.9--49.5&16.6&13.4--20.2&8.4--30.0&15.3&11.4--19.9&7.6--25.8&8.8&6.9--15.0&1.0--22.6\\
SMME&27.0&1.4--37.9&0.5--49.5&15.8&0.7--18.8&-0.1--28.0&14.2&1.1--19.1&0.1--25.8&8.0&0.1--13.2&-0.1--22.6\\
MCPR&28.1&18.1--40.9&14.8--59.0&15.6&10.3--23.7&7.7--33.1&13.7&7.5--21.7&4.8--29.1&7.8&3.2--13.5&0.1--17.6\\
\hline \multicolumn{13}{l}{Hawaii (1981--2010 normal: 2287.7mm)} \\
CMIP5&2.3&-6.1--12.1&-21.0--28.1&-1.3&-11.4--8.4&-17.6--22.7&0.9&-7.7--6.6&-10.8--9.3&3.6&-1.3--10.9&-7.9--21.8\\
SMME&0.5&-4.9--7.8&-20.3--17.9&-0.9&-10.8--6.9&-17.6--13.4&-0.5&-6.1--6.6&-10.8--9.3&0.3&-1.3--8.7&-7.9--12.6\\
MCPR&2.0&-9.9--11.5&-20.2--22.8&-1.0&-9.2--8.8&-21.8--13.2&0.0&-9.7--7.7&-17.0--12.4&2.1&-3.6--10.2&-6.5--16.6\\
\hline
\end{tabular}
\end{table*}


\setcounter{figure}{0}
\renewcommand{\thefigure}{\arabic{figure}}

\begin{figure*}
   \centering
      \includegraphics[width=3.15in]{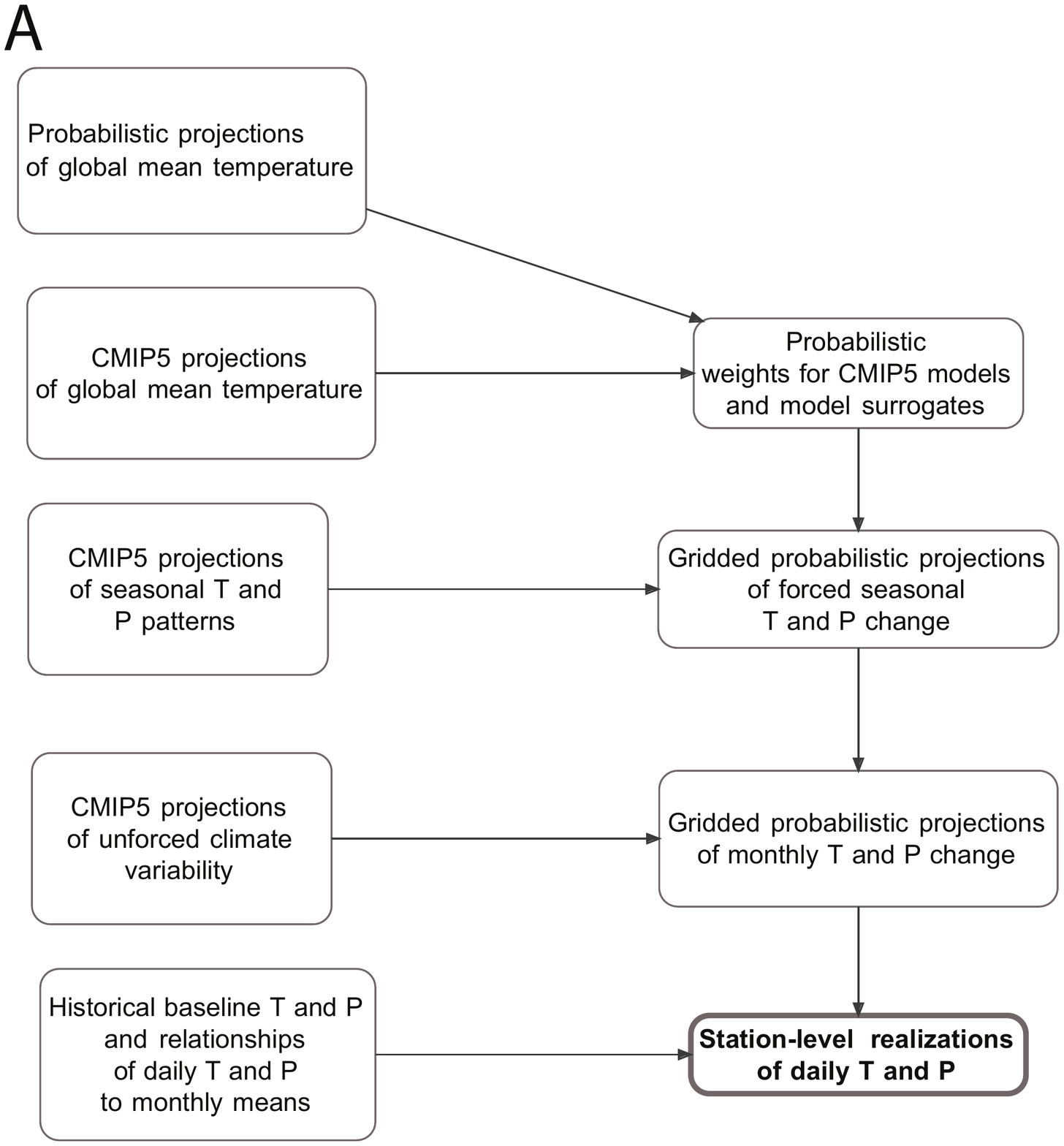}\includegraphics[width=3.15in]{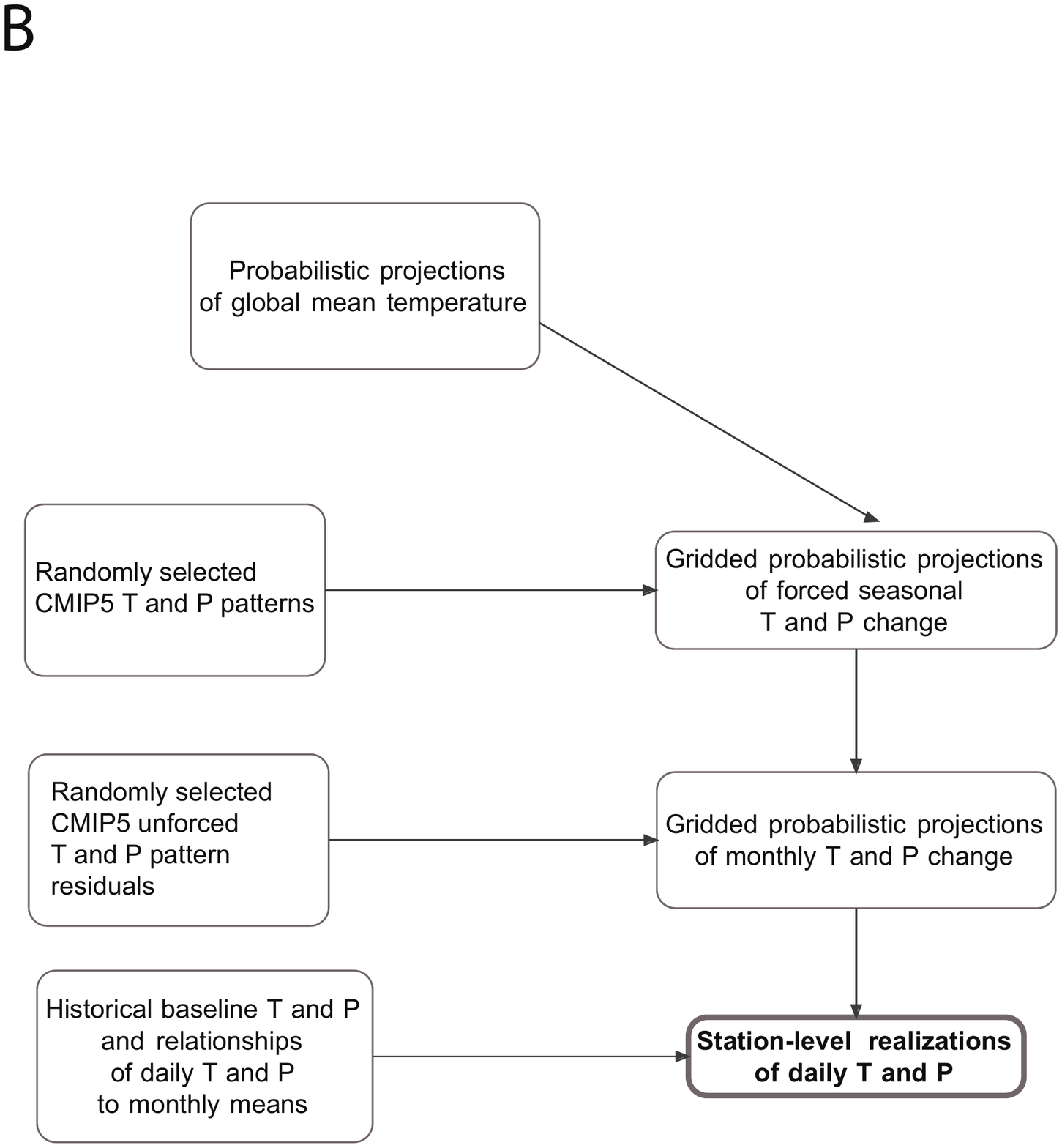}
    \caption{Flow of projection construction for (A) SMME and (B) MCPR.}
   \label{fig:flow}
 \end{figure*}

\begin{figure*}
\centering
\includegraphics[width=0.5\textwidth]{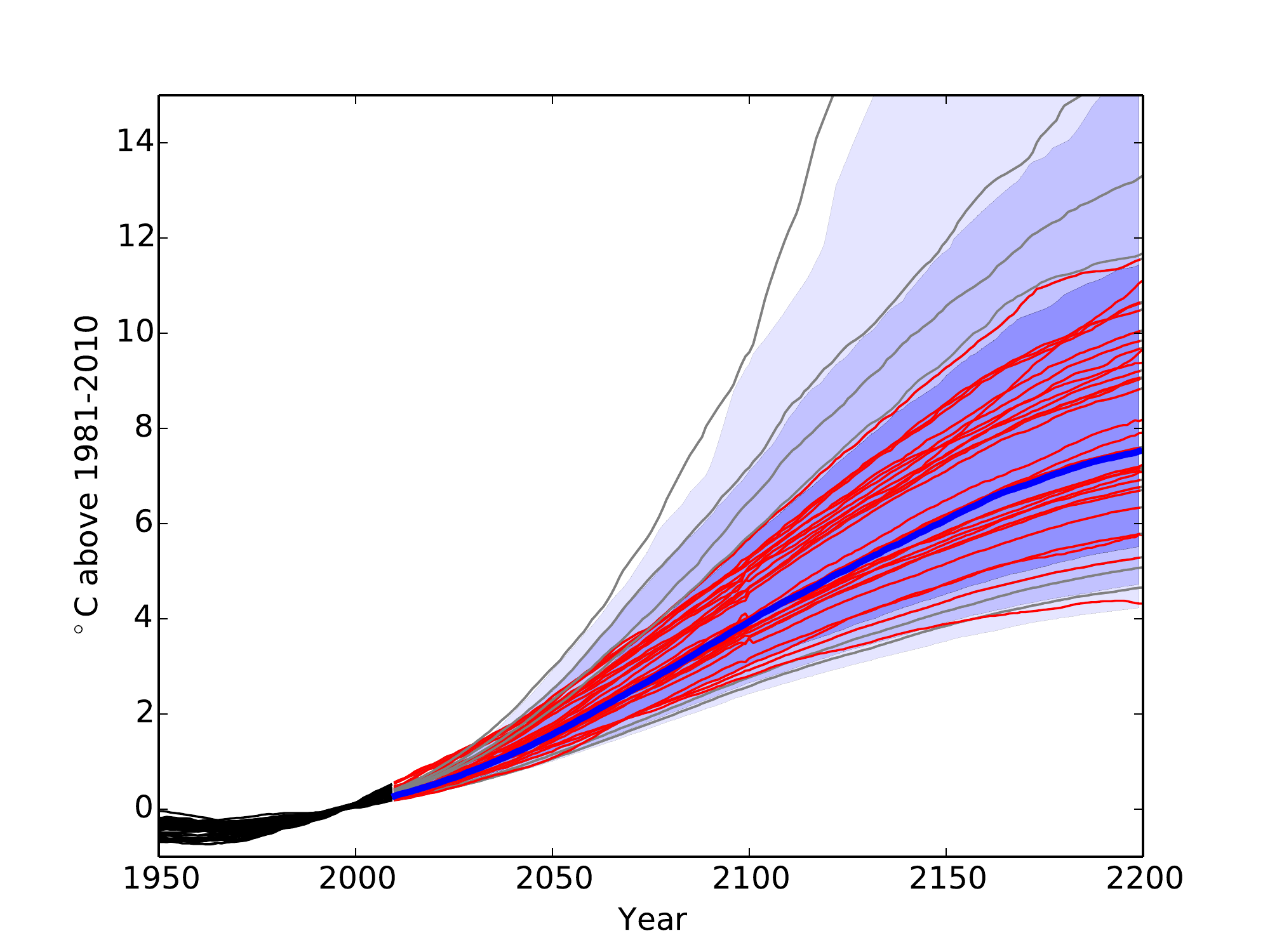}
    \caption{Global mean temperature trajectories for RCP 8.5 from MAGICC6 (thick blue line = median; heavy/medium/light blue shading = 17th--83rd/5th--95th/1st--99th percentiles), individual CMIP5 models (red) and model surrogates used in SMME (grey). }
    \label{fig:global_tas_rcp85}
 \end{figure*}

\begin{figure*}
\centering
      \includegraphics[width=\textwidth]{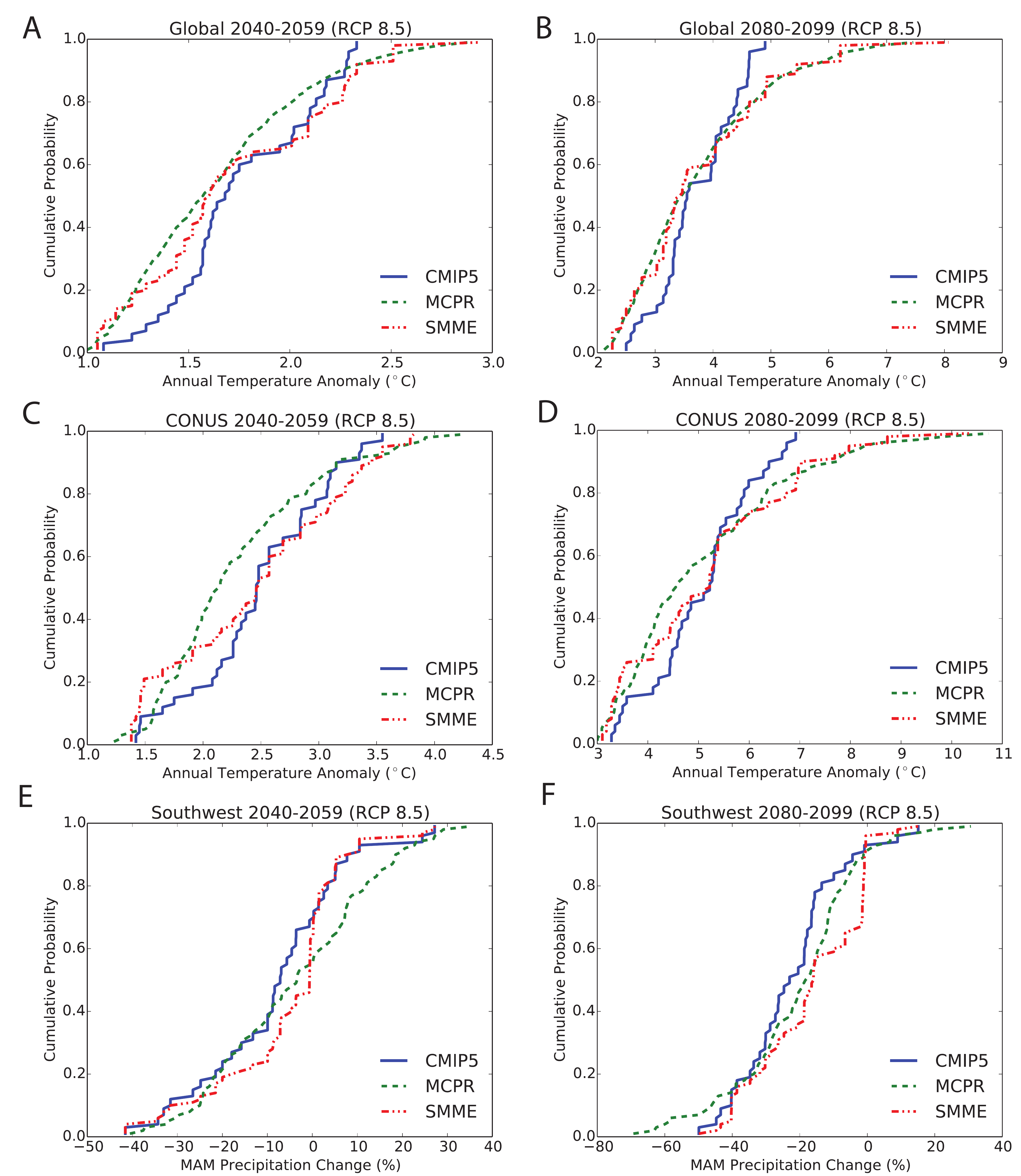} 
    \caption{(A--B) Average (A) 2030--2049 and (B) 2080--2099 global temperature anomaly under RCP 8.5, estimated using the unweighted CMIP5 ensemble (solid blue), MCPR (dashed green), and SMME (dot-dashed red). MAGICC6 global temperature projections are by construction identical to MCPR. (C--D) as for (A--B) but for the contiguous U.S. (E--F) Average precipitation change for the U.S. Southwest under RCP 8.5 for (E) 2030--2049 and (F) 2080--2099.}
\label{fig:tas_prCDF}
\end{figure*}

\begin{figure*}
\centering
      \includegraphics[width=\textwidth]{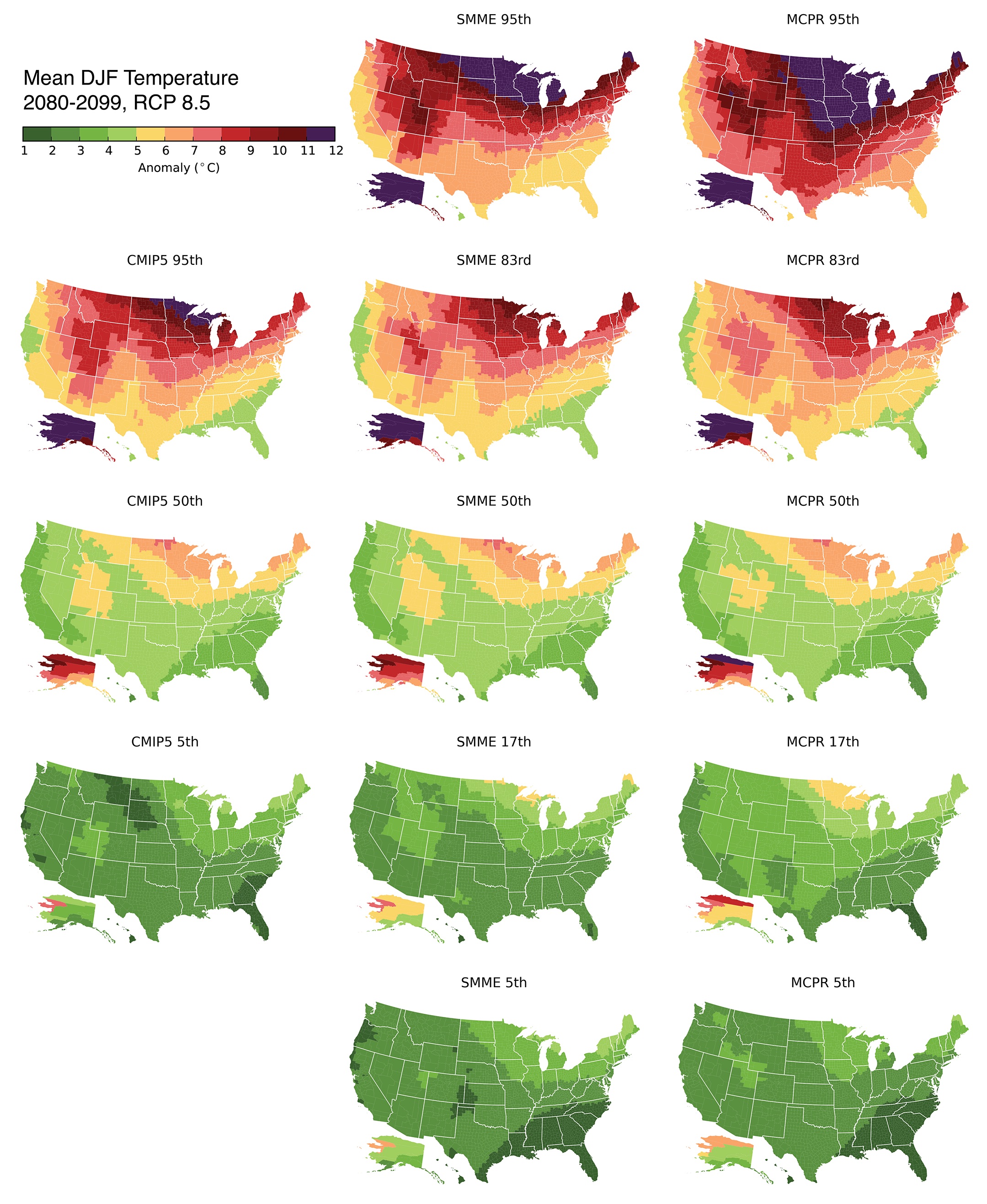} 
    \caption{Average winter (December-January-February) temperature anomaly in 2080--2099 under RCP 8.5 (relative to the 1981--2010 normal) under RCP 8.5 from an equal-weighted CMIP5 ensemble (left) and the SMME (middle) and MCPR probabilistic methods (right). From top to bottom, shown are 95th, 50th and 5th percentiles for the CMIP5 ensemble and 95th, 83rd, 50th, 17th, and 5th percentiles for SMME and MCPR.}
\label{fig:wintertasmap}
\end{figure*}

\begin{figure*}
\centering
      \includegraphics[width=\textwidth]{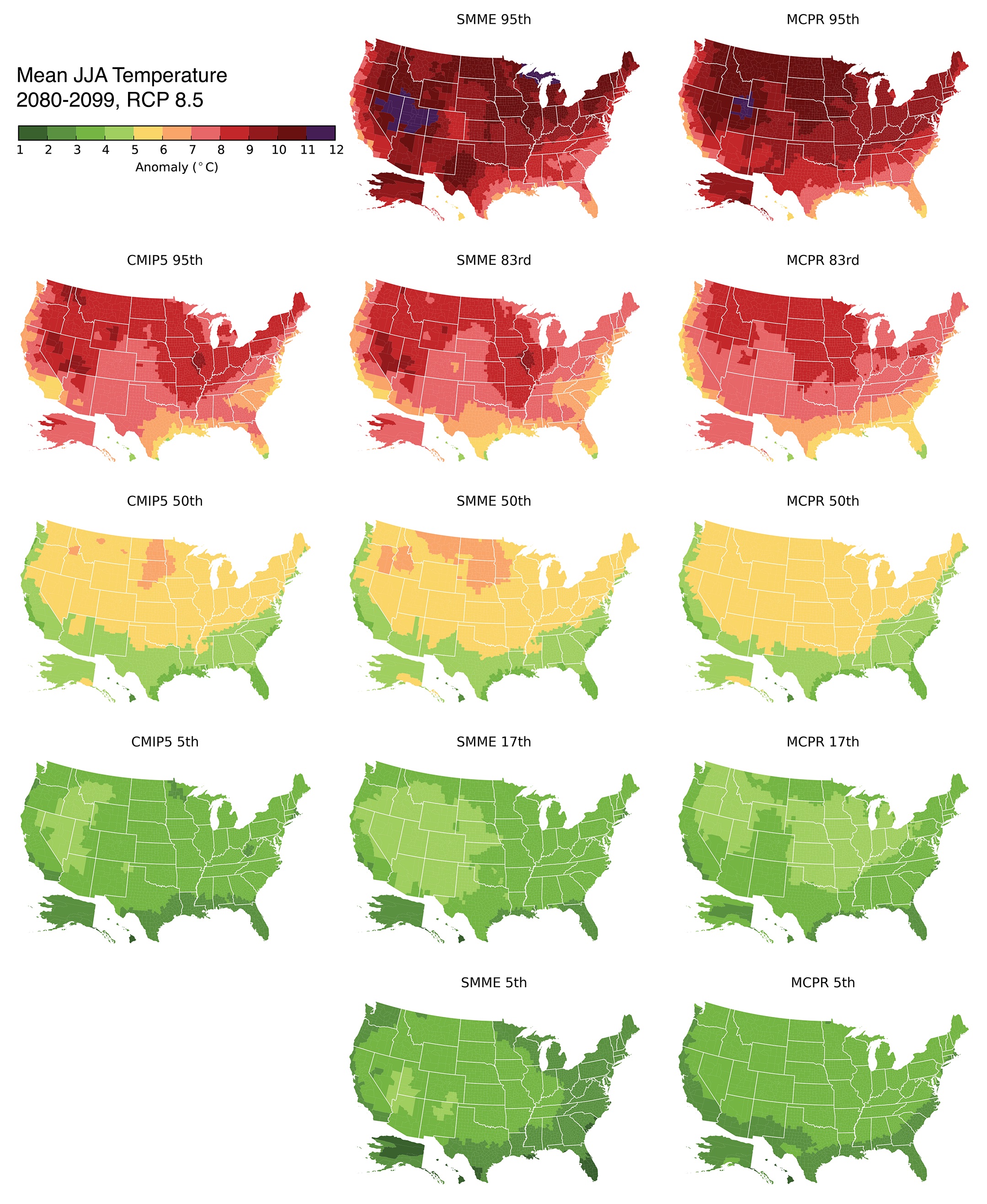} 
    \caption{Average summer (June-July-August) temperature anomaly in 2080--2099 under RCP 8.5 (relative to the 1981--2010 normal) under RCP 8.5 from an equal-weighted CMIP5 ensemble (left) and the SMME (middle) and MCPR probabilistic methods (right). From top to bottom, shown are 95th, 50th and 5th percentiles for the CMIP5 ensemble and 95th, 83rd, 50th, 17th, and 5th percentiles for SMME and MCPR.}
\label{fig:summertasmap}
\end{figure*}

\begin{figure*}
\centering
      \includegraphics[width=\textwidth]{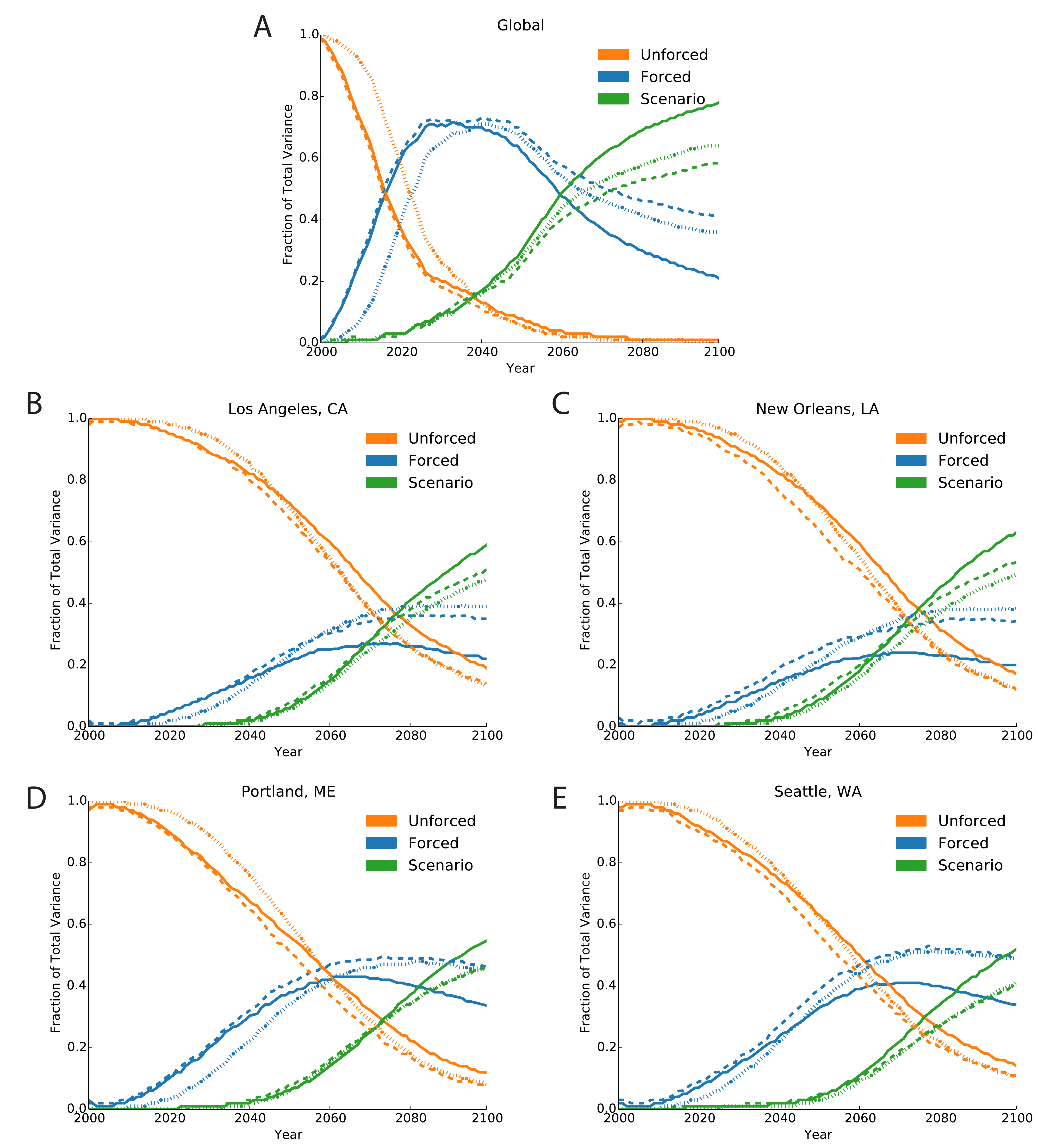} 
    \caption{Fraction of temperature projection variance (solid = CMIP5, dashed = SMME, dotted = MCPR) due to unforced (orange), forced (blue) and scenario (green) uncertainty for (A) global average, (B) Los Angeles, California, (C) New Orleans, Louisiana, (D) Portland, Maine, and (E) Seattle, Washington, temperature}
   \label{fig:variance}
\end{figure*}

\begin{figure*}
\centering
      \includegraphics[width=\textwidth]{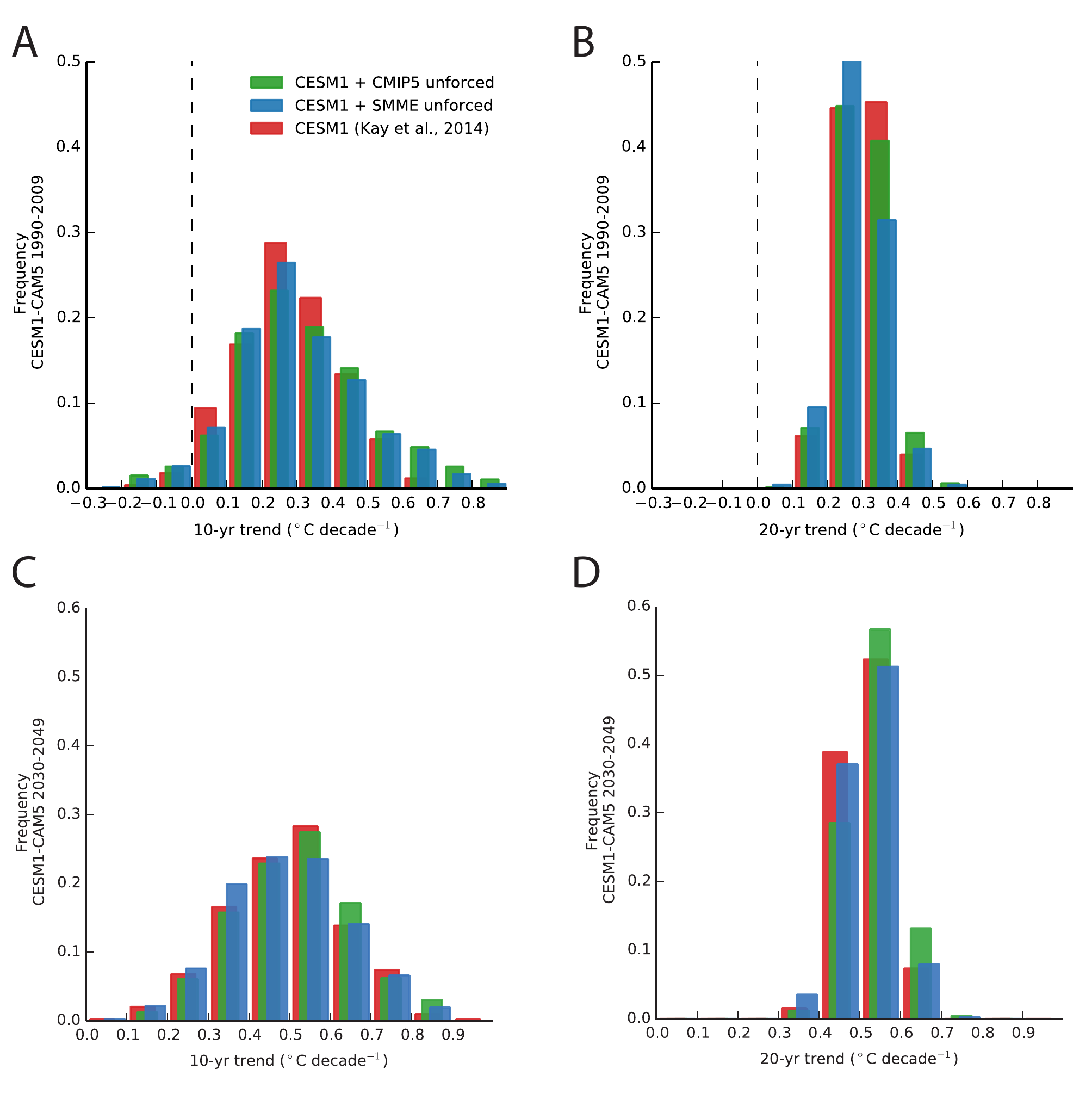} 
    \caption{Distribution of (A) 10-yr  and (B) 20-yr global temperature trends starting from every year between and 1990--2009 using 30 different initializations of the CESM1-CAM5 from Kay et al., 2014 (red), the forced component of the CESM1-CAM5 plus the unforced variability of models in a 33-member CMIP5 ensemble (green), the forced component of the CESM1-CAM5 plus the unforced variability of models from the SMME (blue). (C--D) as for (A--B) but for trends starting from every year between 2030--2049. MCPR randomly samples CMIP5 unforced variability and so yields nearly identical results to the equal-weighted CMIP5 ensemble.}
   \label{fig:kayCMIP5}
\end{figure*}

\end{document}